\documentclass[sigconf, 10pt, nonacm]{acmart}
\author{\normalsize Tella Rajashekhar Reddy$^*$, Atharva Deshmukh$^*$, Karan Tandon, Rohan Gandhi, Anjaly Parayil, Debopam Bhattacherjee}
\thanks{$^*$Equal contribution.}
\affiliation{%
\vspace{0.05in}
  \fontsize{10pt}{12pt}\selectfont{}Microsoft~\country{}
}
\renewcommand{\shortauthors}{Reddy et al.}

\usepackage{tcolorbox}
\usepackage{xspace}
\usepackage{outlines}

\newcommand{\parab}[1]{\vspace{0.03in}\noindent{\bf #1}}
\newcommand{\db}{\textcolor{black}}
\newcommand{\fix}{\textcolor{black}}

\newcommand{\rname}{{\sc beLLMan}\xspace}

\makeatletter
\renewcommand{\@thanks}{\footnote{\@gobbleto@sep\thanks@a}}
\def\thanks@a{\let\@thanks\relax\footnotetext}
\makeatother

\usepackage{graphicx}
\usepackage{subcaption}
\usepackage{float}

\definecolor{lightgray}{gray}{0.92}
\title{\rname{}: Controlling LLM Congestion}

\begin{document}
\sloppy
\begin{abstract}
\db{ 
Large language model (LLM) applications are blindfolded to the infrastructure underneath and generate tokens autoregressively, indifferent to the system load, thus risking inferencing latency inflation and poor user experience.
Our first-cut controller, named \rname{}\footnote{A tribute to Richard E. Bellman\fix{\cite{wikipediaBellman}} for his contributions to control theory.}, enables the LLM infrastructure to actively and progressively signal the first-party LLM application to adjust the output length in response to changing system load. On a real testbed with H$100$ GPUs, \rname{} helps keep \textit{inferencing latency under control} (up to $8\times$ lower end-to-end latency) and reduces energy consumption by $\sim${}$25\%$ (while serving $19\%$ more requests) during periods of congestion for a summarization workload.} 
\end{abstract}

\maketitle

\section{Introduction}

\db{The last few years have seen a tremendous adoption of generative AI, with major infrastructure expansion announcements~\fix{\cite{reuters2025_amazon_ai,2025stargate,reuters2025microsoft,yahoo2025alphabetai,nvidia_xai_ai_infrastructure_2025,Reuters2025AlibabaAI}} from all large AI players. Given that a significant fraction ($\sim${}$90\%$~\fix{\cite{patel2024characterizing, amazon_inference}}) of AI compute today runs AI inferencing, it is imperative that the right interfaces are established between the AI infrastructure and key applications such as large language models (LLMs) so as not to overload the finite resources. Sam Altman, the CEO of OpenAI, had to request users to ``calm down'' and not ``melt'' the GPUs with Ghibli meme generation requests~\fix{\cite{altman2025gpus,reuters2025_ghibli_effect}} reminding us of the congestion collapses~\fix{\cite{CSE461_congestionControl_2017,chaintech_1980_arpanet_crash}} in the early days of the Internet. As LLMs grow from a few billion~\fix{\cite{kamath2025gemma3}} to trillion~\fix{\cite{openai2024gpt4o}} parameters and cater to a global user base, AI providers should have strategies in place that help systems gracefully cope with periods of high load and avoid congestion collapse.}



\db{Currently the LLMs are unaware of the system load and generate tokens auto-regressively to respond to user queries\footnote{We use the terms LLM query/request/prompt interchangeably in this paper.}. Often, LLM responses are verbose, pushing users to rerun queries with additional instructions to shorten the response text. Users also use LLM as a handy tool to summarize long text. This LLM capability to compress text without losing useful information could be leveraged systematically in times of need (high system load).} Note that Claude~\fix{\cite{anthropic2024styles}} offers system prompts that help users set the response style to normal, concise, explanatory, etc.

While one could think of multiple avenues to reduce LLM system load -- using quantized models~\fix{\cite{llama3_8b_quant_sweatycrayfish}}, fewer parameter models~\cite{phi32024}, etc., in this work we explore congestion control using LLM output length reduction. Our idea is backed by the key observations ($O$) and first-cut evaluations ($E$) below:

\parab{($O1$)} LLMs could be concise without noticeably compromising response quality. If not all the time, this interesting property could be leveraged during periods of high system load that increase the response latency. 

\parab{($O2$)} We observe that LLMs, especially the larger ones, have an interesting \textit{emerging capability} -- they closely follow instructions like `write in $X$ words'. Our proposed system, \rname{}, leverages this capability to open a congestion control interface between the system and the application. 

\parab{($E1$)} On a real testbed comprising of a DGX box with $8$ NVIDIA H$100$ GPUs connected via NVLink and having $80$~GB RAM each, we profile a technical paper summarization workload (we pick at random $100$ ACM IMC papers~\fix{\cite{acm_imc}} published between $2022$ \& $2024$) with and without our first-cut \rname{} congestion controller. Our preliminary experiments show how LLM congestion control could effectively keep inferencing latency under control during periods of high load without noticeably affecting output quality.

\parab{($E2$)} \rname{}'s congestion control also reduces the energy consumption by $\sim${}$25\%$ and serve $19\%$ more requests for the workload during congestion -- also demonstrating a sustainability opportunity at scale as a second order effect. 


\section{Background and Motivation}

\begin{figure}[t]
    \centering
    \includegraphics[width=0.7\columnwidth]{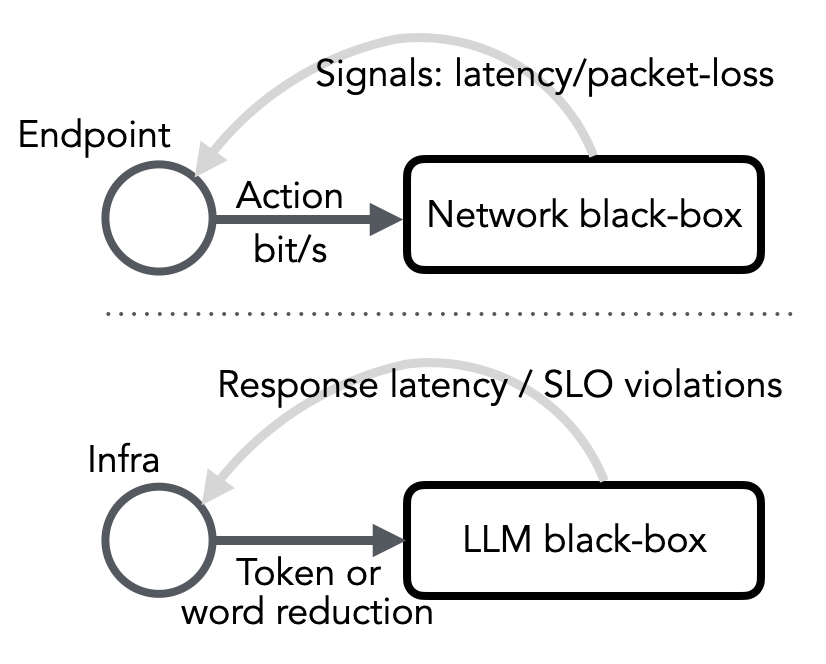}
    \vspace{-0.15in}
    \caption{Internet (top) versus LLM congestion control.}
    \vspace{-0.25in}
    \label{fig:llm_cc}
\end{figure}

In Internet congestion control, endpoints (hosts) react to network load to achieve a better transport experience. The network is considered a black box and the endpoints indirectly measure the network (infrastructure) load by monitoring packet loss~\fix{\cite{tcp_reno}} and/or end-to-end latency~\fix{\cite{tcp_vegas}}. On perceiving high load, the endpoints slow down their sending rates (bits/second), and ramp up otherwise. Fig.~\ref{fig:llm_cc} shows how we could draw parallels with emerging LLM inferencing systems. Although the first-party ($1$P) LLM is still a black-box with the infrastructure being not aware of specific LLM internals, it could still spill out useful signals of load, higher latency or latency SLO (service-level objective) violations for user requests. The LLM system can leverage this load signal to \textit{instruct} the LLM to tone down output generation. In this work, we propose to open a new interface between the system and the LLM that allows the system to implicitly control the volume of tokens (words) generated by the LLM per request in a way that does not compromise on the response quality noticeably while improving query response time during periods of congestion.

\label{subsec:parallels}

\begin{figure*}[ht]
    \centering
    \begin{subfigure}[t]{0.32\textwidth}
        \centering
        \includegraphics[width=\linewidth]{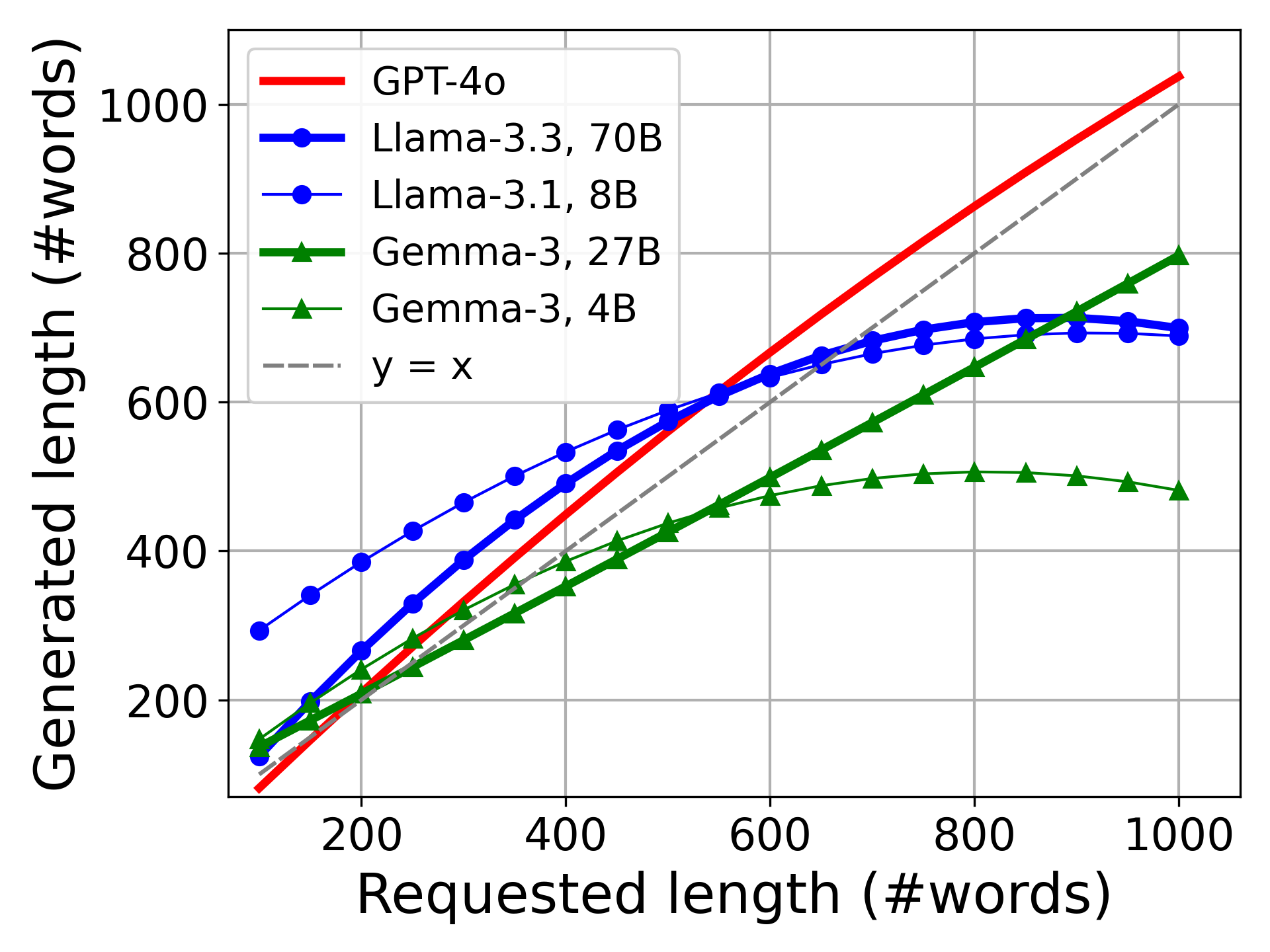}
        \caption{Different Models}
        \label{fig:model_different_models}
    \end{subfigure}
    \hfill
    \begin{subfigure}[t]{0.32\textwidth}
        \centering
        \includegraphics[width=\linewidth]{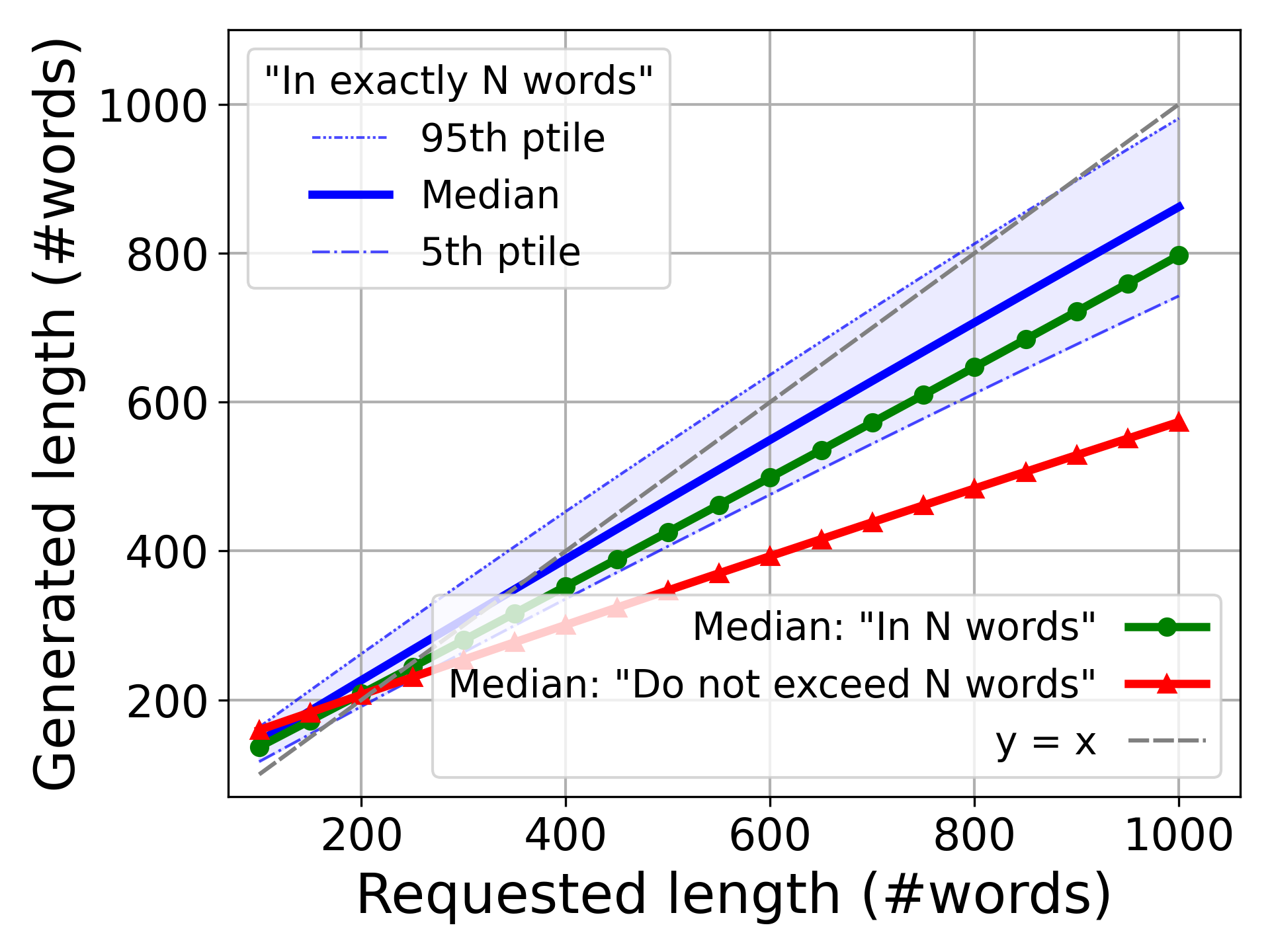}
        \caption{Different Prompts}
        \label{fig:model_different_prompts}
    \end{subfigure}
    \hfill
    \begin{subfigure}[t]{0.32\textwidth}
        \centering
        \includegraphics[width=\linewidth]{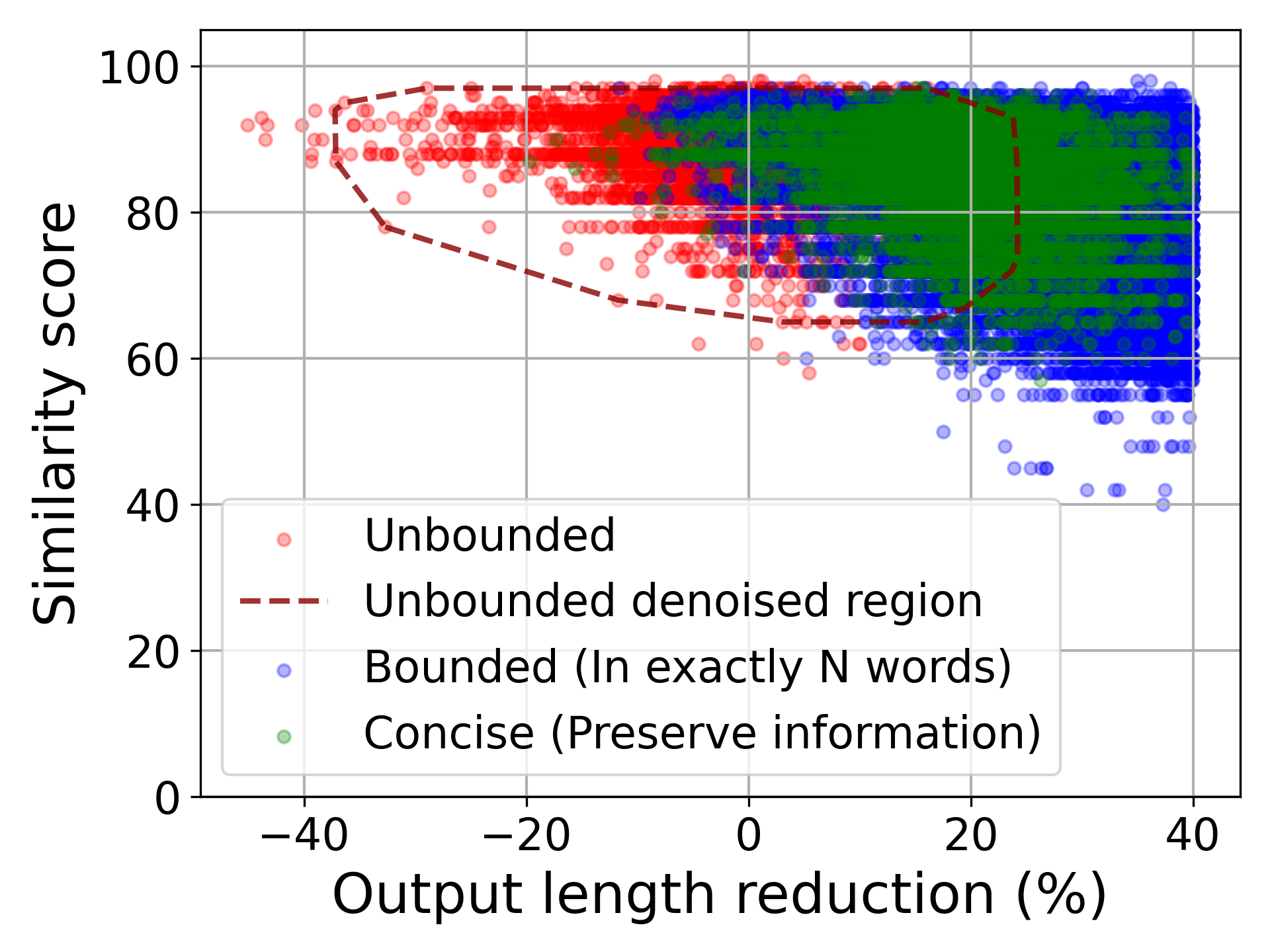}
        \caption{Information Loss}
        \label{fig:info_loss}
    \end{subfigure}
    \vspace{-0.15in}
    \caption{LLM behavior under different configurations.}
    \label{fig:top_three_figures}
    \vspace{-0.15in}
\end{figure*}


\parab{Black-box with implicit signals} In today's real world deployments, the LLM serving systems are unaware of the LLM internals and how they treat individual requests. But schedulers like vLLM~\cite{vllm} have online access to various latency metrics per query -- the queueing latency, the time to first token (TTFT), the time between tokens (TBT), and the E$2$E (end-to-end) latency. As the load on the LLM serving system increases, all these latencies increase, as we discuss in \S\ref{sec:system_results}. 

\parab{The control knobs} During periods of high load, it is plausible to forward the user request to quantized~\fix{\cite{llama3_8b_quant_sweatycrayfish}} or smaller~\fix{\cite{phi32024}} (fewer parameters) models of the same family. Here, we take a different stance and keep using the same model so as not to compromise on the emerging LLM capabilities. Rather, we make an informed choice to reduce the output volume during periods of high load. 
Our work is completely orthogonal to past works~\cite{10.1609/aaai.v39i23.34700,jie-etal-2024-prompt} that fine-tune models to make them understand output length specifications. Previous work~\cite{e3} has addressed the challenge of low GPU utilization in early-exit DNNs by splitting models across multiple GPUs and optimizing batching strategies. In contrast, we propose a congestion control system that can instruct (by prompt appending) such models to generate the query response in $X$ tokens or words instead of the default, predicted $Y$ ($>${}$X$).
In our experiments, we noticed that LLMs from different families (Llama, Gemma, GPT), while internally working with \textit{tokens}, can follow prompt appended instructions significantly better when written in terms of number of \textit{words} as opposed to tokens. 

\vspace{-0.1in}
\subsection{Models listen to us}
To test this hypothesis, we pick models of different sizes and families, and ask to summarize the $100$ IMC papers\footnote{Citations were removed. Both long ($12$ page) and short ($6$ page) papers.}, $10$ time each, in $N$ words, with $N$ varying between $100$ and $1$,$000$ in multiples of $50$. We append a request to each summarization prompt to ``summarize in $N$ words''. For each model and value of $N$, we have $100${}$\times${}$10$ data points (generated output length) and we fit second degree polynomials to the median lengths of the generated outputs (Fig.~\ref{fig:model_different_models}). As we can see, in general, the larger models are closer to the diagonal ($y=x$, $x$ being the target length in the prompt and $y$ being the generated output length) for the same family, and across models GPT-$4$o~\fix{\cite{openai2024gpt4o}} (projected $1.7$T parameters~\fix{\cite{cognitivetech2025_gpt4_params}}) is closest to the diagonal, followed by Llama-$3.3$ ($70$B)~\fix{\cite{meta2024llama3.3_70b}} and Gemma-$3$ ($27$B)~\fix{\cite{kamath2025gemma3}}. This demonstrates an emerging capability -- the larger models are able to stick to word limits better. Although there is still scope for perfect alignment with the diagonal, the strong trends validate our hypothesis.

\parab{Scope for prompt engineering} Next, we pick a specific model, Gemma-$3$ ($27$B), with an aim to bring the bounded generations closer to the diagonal ($y=x$) simply by improving the prompt. We see in Fig.~\ref{fig:model_different_prompts} that out of the $3$ prompts that we tried, ``summarize in exactly $N$ words'' works best. It is quite interesting to see how adding the keyword ``exactly'' significantly improves the model's alignment to the prompt instructions. 
The shaded region in Fig.~\ref{fig:model_different_prompts} shows that most of the generations are close to the diagonal. For the rest of this work, we use Gemma-$3$ ($27$B) as our test LLM and the best prompt here as the interface between the system and the application. 

We are optimistic that this emerging capability could improve further, as is evident from the trends. GPT-$4$o is already very close to the diagonal as seen in Fig.~\ref{fig:model_different_models}.

\vspace{-0.15in}
\subsection{Taming the LLM uncertainty}

\parab{Uncertainty by default} LLMs are uncertain in generating output due to their probabilistic and auto-regressive nature. Even with \texttt{temperature}~\fix{\cite{murel2024llm_temperature}} set to $0$ and \texttt{top\_p}~\fix{\cite{ibm_watsonx_prompt_params}} set to $1$, there is still some residual uncertainty left~\fix{\cite{atil2025nondeterminismdeterministicllmsettings}}. During conversations with LLM experts we learned that the \texttt{temperature} is often set to $0.3$ to avoid hallucinations and generate consistent output~\fix{\cite{promptengineering_temperature_top_p}}
, and for chatbot or creative writing, the values are often higher. For our work, we stick to the value of $0.3$ and quantify the default uncertainty in LLM generation. We generate $100$ summaries for each of the $100$ IMC papers with Gemma-$3$ ($27$B), and for each paper pick the summary with the median (out of $100$) output length as the reference point. Fig.~\ref{fig:info_loss} shows (in red, labeled as `unbounded') the variations in output lengths (percentage differences with respect to the reference points) for all papers and all summaries. On the $x$-axis, this default variation mostly lies between $\sim${}$-38\%$ (inflation in size) and $\sim${}$25\%$ (the dashed border devoid of the $0.1\%$ tail) showing how the output length could vary significantly between multiple runs of the same query. So, a simple intuition we have is that LLM users might not even perceive a calculated reduction in output length (well within a safe bound of $20\%$) if the output quality is \textit{on par}. However, 
we should use this knob with caution, as we discuss below, to not affect user experience to a noticeable degree.

\parab{Understanding output quality} It is crucial to identify an output quality check metric to not noticeably compromise the default user experience even during times of high system load.
Lexical metrics such as ROUGE score~\fix{\cite{lin2004rouge}}, entropy~\fix{\cite{entropy}}, and the matching of named entities~\fix{\cite{nasar2021named}} do not work well due to the lack of understanding of the semantics of text, the nuances of natural language and context. Hence, we use LLM as a judge~\fix{\cite{li2024generation,gu2024survey,wei2024systematic,chiang2023largelanguagemodelsalternative}} to measure the similarity scores,  information loss percentage, between pieces of text ($2$ different summaries -- reference and test). In our exhaustive unit tests (selectively removing units of information and sentences from the test text), we observed that OpenAI \texttt{o$3$} (reasoning model)~\fix{\cite{openai2024o3}} is consistent and reasonable in assigning similarity scores to pairs of text and performs significantly better than GPT-$4$o. We made sure that the content similarity scores (prompt in Appendix~\ref{appendix:eval_prompt}) reflect the loss of information content and the loss of entropy but do not get skewed by the core theme which is expected to be similar in this context.
In Fig.~\ref{fig:info_loss}, we see that even in the unbounded case, the similarity scores vary significantly with respect to the reference point (median length, as discussed above). While toward the left of the unbounded denoised region (larger sizes) the scores are higher, toward the right the similarity scores are relatively low. The takeaway is that, even without any restrictions, the similarity scores could be as low as $65\%$, offering us a window of opportunity to reduce LLM output lengths during high system load while still meeting this bar.

\parab{The rigid `concise'} Before we quantify this opportunity window more systematically, we explore a quick hack -- can \rname{} controller use a boolean knob and ask the LLM to be `concise' by appending to the prompt during periods of load? The green region in Fig.~\ref{fig:info_loss} represents such generations for the same workload, and similarity scores and output length differences measured with respect to the unbounded generation with median output length. While the lowest similarity scores are still on par with the unbounded generations, the output length reductions could be significant. One might think this is a great choice, but here is a word of caution: this knob is pretty rigid, and we cannot force it to operate close to the top left corner of Fig.~\ref{fig:info_loss} even during a mild system load. In such scenarios, since we lose control over the output length, the similarity scores could be affected for no reason.

\parab{Output length versus quality} As models listen to prompt instructions to ``summarize in exactly $N$ words'', we generated summaries with $x\%$ reduction in output length (blue region in Fig.~\ref{fig:info_loss}) for $x=2,4,...,8,10,15,20,...,40$ with respect to the reference unbounded generation (median length). This bounded region gives us an operational window of $<${}$20\%$ (we deliberately keep it well below the $24\%$ opportunity for safety) output length reduction, with similarity scores on par with unbounded generation. Note that during periods of mild load, the requested output length reductions could be kept low. This is part of the `minimally invasive' strategy we intend to pursue to avoid unintended user experience issues.

\parab{Leveraging output length predictors} Output length predictors are widely adopted 
for scheduling\fix{~\cite{dynamo,elis, aladdin, userve, zheng2023responselengthperceptionsequence}}, routing\fix{~\cite{jain2025intelligentrouterllmworkloads,s3}}, and energy-aware routing \fix{~\cite{reddy2025aigreenferencingroutingai}}. When an LLM request arrives online at the queue, it is important to predict (best effort) the unbounded output length to be able to right-size the bounded output depending on system load. Toward this, we can leverage the output of any existing prediction pipeline for scheduling and routing. In our setup, it consists of a Longformer\footnote{Works best in comparison to ModernBERT~\fix{\cite{modernbert}}, Jina~\fix{\cite{jinaembeddingsv3}}, and BGE-m$3$~\fix{\cite{bgem3}}.} encoder model~\fix{\cite{beltagy2020longformerlongdocumenttransformer}} that produces embeddings from the prompt text, which are then passed through a two-layer feedforward neural network followed by a final layer that gives the predicted output length in number of words. The predictor was trained and tested ($80$:$20$ split) on $\sim${}$11$,$000$ technical papers from ICLR ($2020$-$23$)~\fix{\cite{iclr_about}} downloaded with OpenReview API~\fix{\cite{openreviewpy}}. For the training set, $5$ summaries were generated per paper with Gemma-$3$ ($27$B). The predictor achieved an MAE (Mean Absolute Error) of $36$ and an inference latency of $50$~milliseconds on a single H$100$ GPU. Note that this prediction is not on the critical path and is triggered only during periods of high load while the requests are still in the queue. 
\section{Our idea \& initial results}
\label{sec:system_results}

\begin{figure}[t]
    \centering
    \includegraphics[width=0.85\columnwidth]{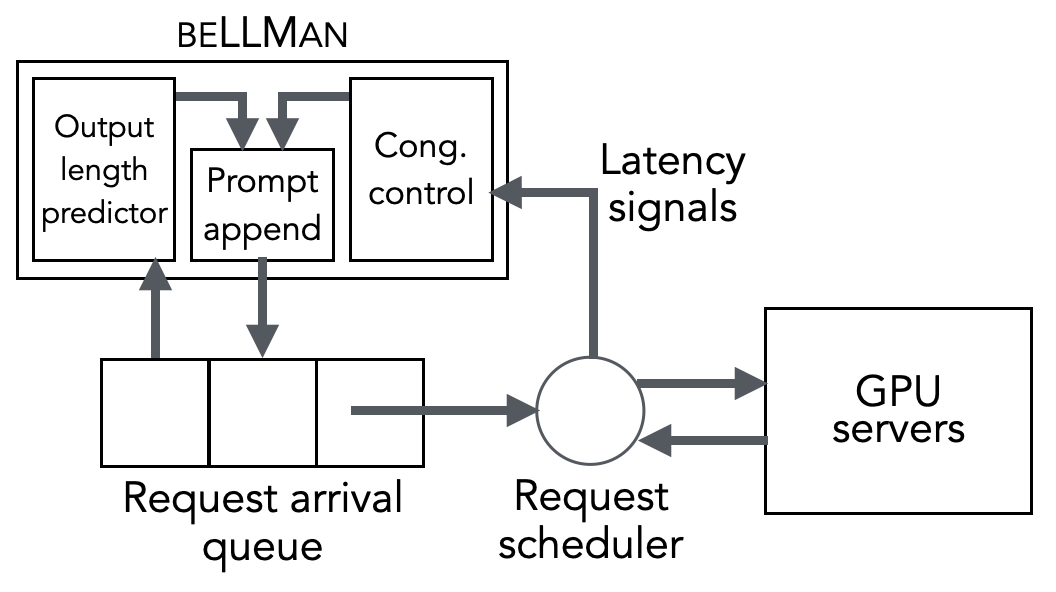}
    \vspace{-0.15in}
    \caption{\rname{} architecture diagram.}
    \vspace{-0.25in}
    \label{fig:architecture}
\end{figure}

\db{Our key proposal is \rname{} -- a $1$P LLM serving system component, as in Fig.~\ref{fig:architecture}, that ($1$) works with real-time latency signals from the LLM request scheduler, ($2$) manipulates LLM requests in the arrival queue, and ($3$) thus instructs the LLM application to generate shorter output during high system load. In an LLM serving system, the request scheduler (we use vLLM scheduler) picks requests from the arrival queue and assigns to GPU servers. The scheduler keeps track of request-level metrics like queueing latency, TTFT, TBT, and end-to-end latency (latency components discussed in \S\ref{subsec:parallels}). \rname{} consumes these latency signals from the scheduler. When \rname{} senses congestion (latency inflating over time), it triggers its congestion control algorithm to decide the word reduction rate ($r$) between $5\%$ and $20\%$ (so as not to significantly affect the LLM response quality). As individual LLM requests are queued, \rname{} predicts the unbounded output length ($L$), calculates the reduced length ($N=L${}$\times${}$(1-r)$), and appends an instruction to the prompt to generate summary `in exactly $N$ words'. This way, the output length prediction overhead is hidden behind the request queueing latency. 
\rname{} complements existing mitigation techniques (quantized or smaller models) while avoiding model swapping.
Note that the word reduction is enforced only when latency inflates due to congestion.} 

\parab{Placeholder for congestion control} \rname{} currently runs a simple, first-cut linear congestion control while offering quick and easy integration of more complex schemes in the future. When congestion is sensed (TBT starts to inflate), it triggers a reduction in output length with $r=5\%$. The $r$ values increase linearly until $20\%$ if the maximum latency threshold is reached. Also, $r$ is reset when the TBT latencies are low. We discuss these thresholds below in the context of the specific trace on which we test \rname{}.


\begin{figure*}[ht]
    \centering
    \begin{subfigure}[t]{0.33\textwidth}
        \centering
        \includegraphics[width=\linewidth]{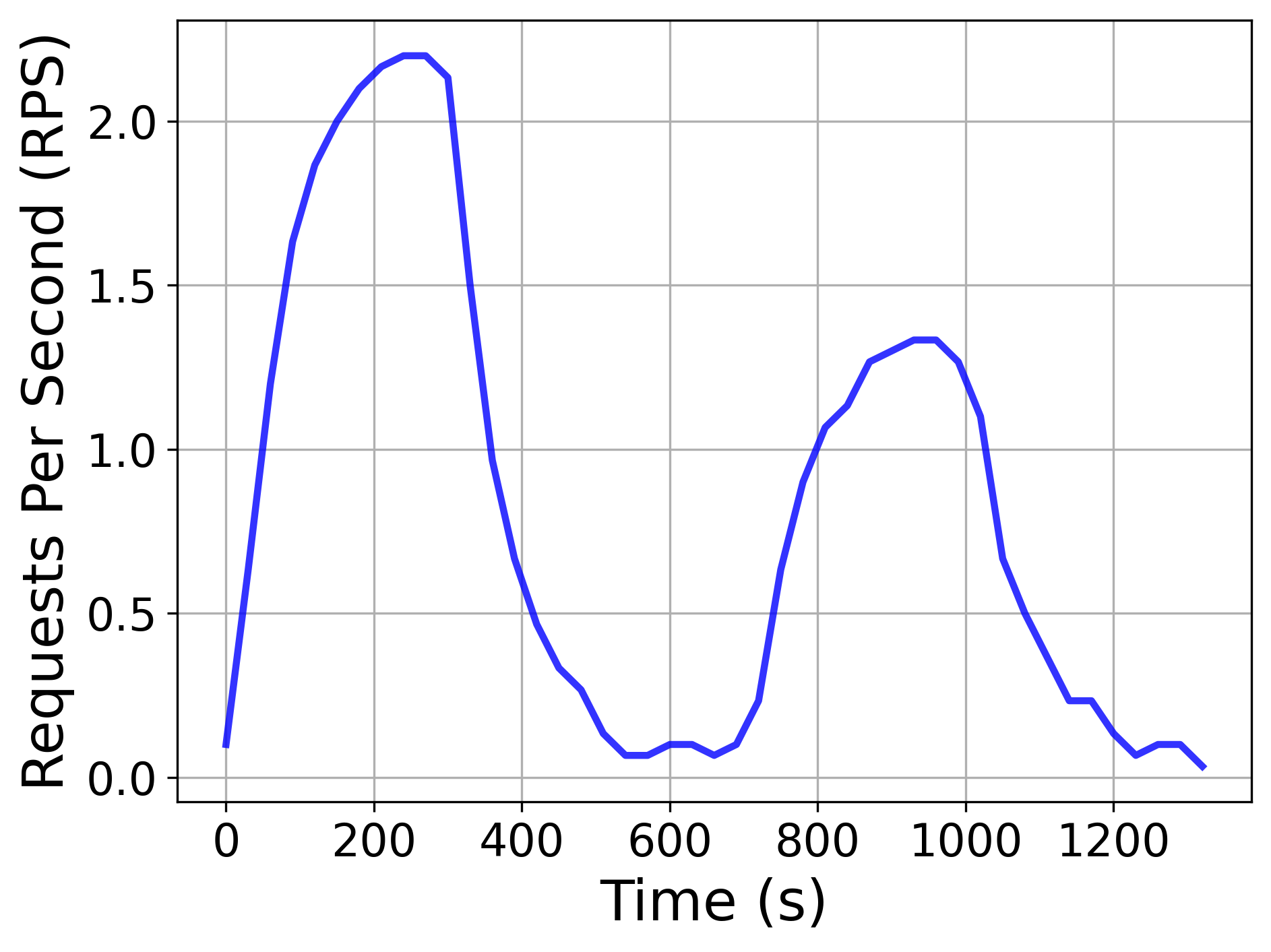}
        \caption{Paper summary trace}
        \label{fig:trace}
    \end{subfigure}
    \hfill
    \begin{subfigure}[t]{0.33\textwidth}
        \centering
        \includegraphics[width=\linewidth]{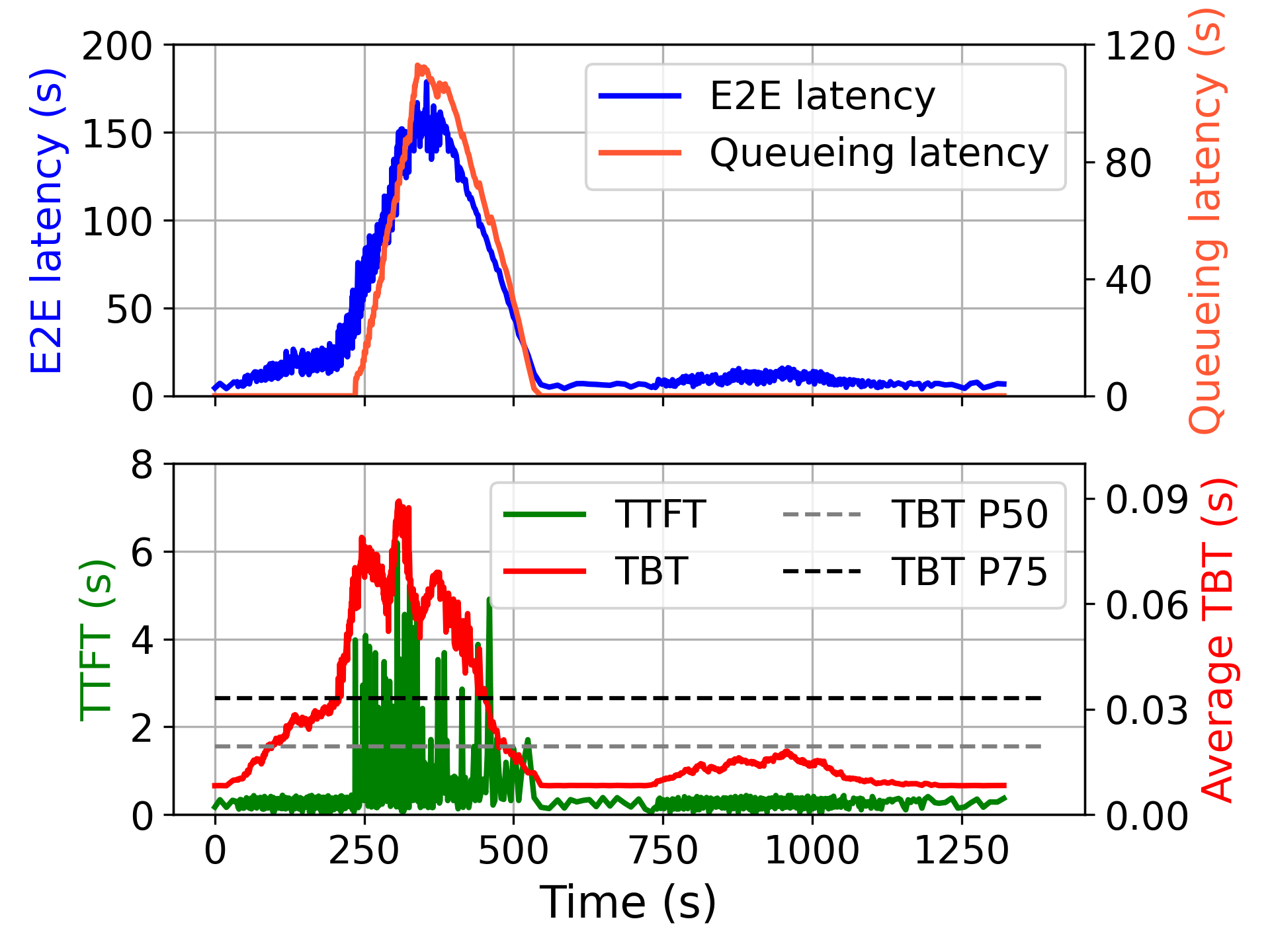}
        \caption{Unbounded run (default)}
        \label{fig:unbounded_run}
    \end{subfigure}
    \hfill
    \begin{subfigure}[t]{0.33\textwidth}
        \centering
        \includegraphics[width=\linewidth]{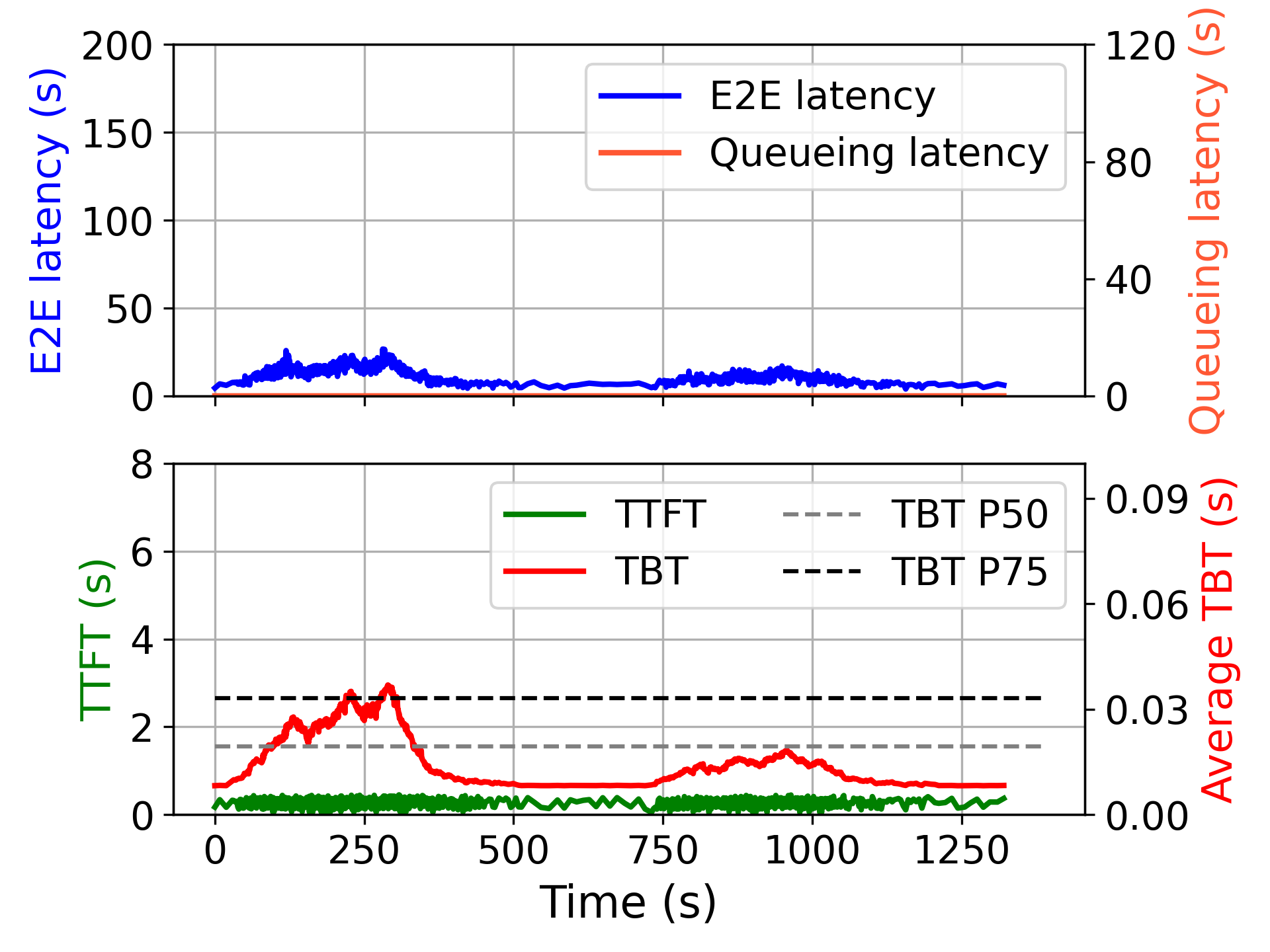}
        \caption{Bounded run with \rname{}}
        \label{fig:bounded_run}
    \end{subfigure}
    \vspace{-0.15in}
    \caption{Synthetic trace and the corresponding unbounded and bounded runs.}
    \vspace{-0.15in}
    \label{fig:cc_results}
\end{figure*}

\parab{Setup and trace} \db{For all experiments below, including trace generation, we use the same LLM and setup: Gemma-$3$ ($27$B) LLM hosted on a DGX box with $8$ NVIDIA H$100$ GPUs connected via NVLink and having $80$~GB RAM each, configured with tensor parallelism of degree $8$ (TP$8$). We use the vLLM scheduler (with the only modification to stream latency signals to \rname{}) and disable prefix-caching to avoid contaminating our observations. For trace generation, first we run the paper summary workload (papers picked uniformly at random from the $100$ IMC papers with duplicates allowed) on the above setup. We test for different requests per second (RPS) values starting at $1$, for $10$ min each, and observe queueing latency linearly increasing over time (for a particular run) starting at an RPS value of $2.4$ indicating significant overload. With this observation in mind, we generate a synthetic trace using the Poisson process, where inter-arrival times follow an exponential distribution -- similar to prior works~\cite{li2023alpaservestatisticalmultiplexingmodel,jiang2024hexgengenerativeinferencelarge}. The trace spans a duration of $22$ minutes with $2$ peaks of $2.5$ RPS (high load) and $1.5$ RPS (parameters still under check; no need to trigger congestion control), as shown in Fig.~\ref{fig:trace}. The trace has distinct phases when the request arrivals ramp up, stay put, and ramp down, and the target RPS values are achieved with Poisson arrivals.}

\parab{Default unbounded run} \db{During the unbounded run with no congestion control in place, all latency signals spike, as observed in Fig.~\ref{fig:unbounded_run}, during the first RPS peak of $2.5$ that lasts for $\sim${}$90$ seconds. The queueing latency shoots up quickly as compute saturates and incoming requests have to wait. This high queueing latency also reflects in the E$2$E latency inflation. As GPUs go full-throttle to consume and spill out more tokens (words) in aggregate, the TTFT latencies also increase. Note that TBT latency inflation occurs sooner as incoming RPS values go up over time, and hence could offer early signals of an impending congestion. Based on this key observation, we use TBT as the congestion signal for the rest of this work. We pick the median TBT ($T1$) during the unbounded run as the threshold for triggering congestion control ($r=5\%$) in the bounded run with \rname{} and the $75^{th}$ percentile TBT ($T2$) as the (soft) maximum acceptable TBT threshold that corresponds to $r=20\%$ in the bounded run (dashed horizontal lines in Fig.~\ref{fig:unbounded_run} and \ref{fig:bounded_run}). The second shorter peak of $1.5$ RPS in the trace does not see any major congestion with slight inflation in TBT and E$2$E latency.}

\parab{\rname{} in action} \db{\rname{} consumes the average TBT each second from the vLLM scheduler and computes a moving average over the last $5$ seconds. If the moving average (to avoid unnecessary triggers in haste) exceeds $T1$ (from the unbounded run), it triggers congestion control with $r=5\%$. It linearly increases $r$ until the threshold $T2$ is reached (if at all) when $r=20\%$. For the specific trace, \rname{} triggers congestion control at $131$s and stops at $351$s (based on the TBT moving average) significantly sooner than when congestion ends in the unbounded (default) case as things are under control. When congestion control is active, the median $r$ across all requests is $8\%$. TBT remains roughly within bounds, queueing latency and TTFT are kept low, and E$2$E latency is reduced by up to $8${}$\times$. \rname{} could not only improve the latency components but also the duration of congestion at the cost of very modest drop in quality during congestion. We compared the similarity scores for individual LLM requests during the bounded run (with respect to the unbounded run) while congestion control is active and inactive (same as unbounded). The median similarity scores with respect to the unbounded run are $87\%$ when congestion control is active and $88\%$ when inactive. This observation confirms that the drop in output quality with our controlled output length reduction during congestion is minor.}

\parab{\rname{} energy savings} \db{During the congestion period ($130$-$500$s) in the unbounded run, \rname{} helps consume $25\%$ less energy (as reported by the vLLM scheduler) and drives $19\%$ more LLM requests to completion thanks to output length reduction.  This observation also opens up another possibility of using \rname{} to reduce the energy consumption of AI/LLM systems at scale. Note that the energy savings are due to shorter outputs and not improved GPU efficiency.}


\parab{Additional observations} \db{We tested \rname{} on a trace with the $2.5$ RPS peak lasting $2${}$\times$ longer. We still did not observe any major queueing latency. On increasing the peak from $2.5$ to $3.5$ RPS in the trace, \rname{} experiences some queueing latency (tipping point for finite hardware capability) but all latency components are significantly lower than in the unbounded case.}
\vspace{0.1em}

\vspace{-0.2in}
\section{Discussions and Future Work}

\parab{LLM congestion signals} While we focus on TBT as a signal for congestion, other latency components could also be relevant. While E$2$E latency signals are delayed, TTFT could offer early insights. An even earlier signal could be the input tokens per unit time to the LLM serving system. GPU utilization metrics can also serve as useful signals.

\parab{Toward novel LLM congestion control} An \textbf{MPC} (model predictive control) controller~\fix{\cite{wikipediaMPC}} might be able to predict the system load and preemptively adjust $r$ better than the current reactive linear controller. It should be able to use cost functions that include latency, output quality, and energy efficiency over a moving time horizon and help avoid oscillations in $r$. A \textbf{BBR}~\fix{\cite{cardwell2016bbr}} style controller could proactively operate at the equivalent of Kleinrock's point~\fix{\cite{kleinrock1979power}}, keeping TBT low while maximizing the token generation (tokens/s) of the serving system. A \textbf{PCC}~\fix{\cite{dong2015pcc}} controller could run micro-experiments with different values of $r$ and measure a utility function that combines both latency and response quality.


\parab{Nuanced fairness issues} Fairness across flows~\cite{CHIU19891,jacobson1988congestion} is key in Internet congestion control and applies to LLMs too -- both per-request and per-class (priority). For requests, using the same $r$-value within a batch ensures equal treatment. For classes, $r$ can vary; e.g., coding tasks might use $r=0$ to avoid errors, and very short outputs can bypass size reduction. A logically centralized \rname{} system could help experiment with different aspects of fairness with tractable insights.

\parab{Broad use across applications} \db{In this work, we focus on a document summarization workload. This is just one of many AI/LLM inferencing use cases. Although in certain cases, such as coding, it is not practical to even try such compression, a large fraction of LLM use cases like conversation, synthesis, etc. can benefit from integrating \rname{}.}

\parab{Prediction-less design} While output length prediction is a low overhead task with inferencing latency hidden behind request queueing, some operators might not want this component in their design. In such cases, different prompts could be potentially used as proxies for different degrees of reduction (Fig.~\ref{fig:model_different_prompts}) based on the system load. There is scope for more fine-grained prompt engineering.

\vspace{-0.15in}
\section{Conclusion} In this work, we propose \rname{}, which offers a framework to plug in congestion control to LLM serving systems. Our first-cut experiments show that an interface between the system and the LLM application actually works, keeping serving latencies under check during periods of increased load at the cost of modest drop in output quality.




\bibliographystyle{ACM-Reference-Format} 
\bibliography{faisys}

\appendix
\section{Evaluation Prompt}
\label{appendix:eval_prompt}
In our experiments, we tried various OpenAI \texttt{o$3$} (reasoning model)~\fix{\cite{openai2024o3}} prompt formulations to evaluate the similarity between texts and picked the following that is the most consistent. 

\begin{tcolorbox}[colback=gray!10, colframe=gray!80, title=Evaluation Prompt]
Give me a similarity score between 0 (no similarity) and 100 (exactly similar) for the test text compared to the reference text below. The score should reflect the information content loss and entropy loss in the test text compared to the reference text. The score should not reflect the core theme or the main idea of the text which is same anyway. Think and then arrive at a single score.

Reference text: \texttt{<text>}

Test text: \texttt{<text>}
\end{tcolorbox}

\end{document}